\documentclass[preprint,aps]{revtex4}
\usepackage{amsmath}
\usepackage{graphics}
\usepackage{epsfig}
\usepackage{subfigure}
\usepackage{fancyheadings}
\usepackage{tabularx}
\usepackage{amssymb}
\usepackage{oldgerm}
\def\be{\begin{equation}}
\def\ee{\end{equation}}
\begin{document}
\title{Pairing in 4-component fermion systems: the bulk limit of
$SU(4)$-symmetric Hamiltonians}
\author{G. F. Bertsch$^1$}
\author{J. Dukelsky$^2$}
\author{B. Errea$^2$}
\author{C. Esebbag$^3$}
\affiliation{$^1$Institute for Nuclear Theory and Dept. of Physics,
University of Washington, Seattle, Washington.\\$^2$Instituto de Estructura de la Materia - CSIC, Serrano
123, 28006 Madrid, Spain. \\$^3$ Departamento de Matem\'aticas, Universidad de Alcal\'a, 28871
Alcal\'a de Henares, Spain.}

\begin{abstract}
  Fermion systems with more than two components can exhibit pairing
condensates of much more complex structure than the well-known single BCS
condensate of spin-up and spin-down fermions.  In the framework of
the exactly solvable $SO(8)$ Richardson-Gaudin (RG)  model with $SU(4)$-symmetric Hamiltonians, we show
that the BCS approximation remains valid in the
thermodynamic limit of large systems for describing
the ground state energy and the canonical and quasiparticle excitation gaps.
Correlations beyond BCS pairing give rise to a spectrum of
collective excitations, but these do not affect the bulk energy and
quasiparticle gaps.
\end{abstract}
\maketitle
\section{Introduction}
The exact solutions of the pairing Hamiltonian with complete orbital
degeneracy are available as simple analytic formulas not only for
ordinary $S=0$ pairing of spin $1/2$ particles with $SU(2)$ symmetry \cite{RiSch}, but also for
systems of particles having additional degrees of freedom with respect to
an internal quantum number, e.g. color degeneracy in quark models \cite{Petry85,Bohr09} with $SO(6)$ symmetry, or
the isospin degeneracy of neutrons and protons in atomic nuclei
\cite{fl64,ev81} with $SO(8)$ symmetry.
    To fully describe pairing in nuclei, we
should consider an isospin-invariant pairing Hamiltonian
made of the three $S=0$ pair creation and annihilation
operators transforming under isospin as $T=1$ closing an $SO(5)$ algebra.  We call this the isovector
pairing Hamiltonian.  In nuclear systems where the number of neutrons $N$ is close to the number of protons $Z$, we should also consider the $T=0$ pairing
leading to a Hamiltonian made of the six pair operators that
carry $S=0,T=1$ and $S=1,T=0$.  We show that the bulk limit of the
pair correlation energy in the systems with neutrons and protons is
equal to sum of the pairing energies of the two kinds of nucleons
in isolation.  Also, the gap in quasiparticle excitation energies is the
same.  This demonstration supports the use of the BCS theory
to calculate pairing energies for the bulk limit of fermion systems with
more than two degenerate degrees of freedom.

The exactly solvable pairing Hamiltonians are all based on interactions
that have a simple operator structure.  The first we consider
is the pure seniority $SO(8)$ Hamiltonian of $T=0,1$ pairing, for which analytic expressions
for the energy can be derived.  This model has been applied to nuclear
physics in ref. \cite{en97}.  More realistic but still
exactly solvable by algebraic methods are the RG Hamiltonians with include non-degenerate single particle levels (for a review see \cite{du04}).
More recently, these exactly solvable models have been extended to higher rank algebras \cite{duk06,duk09,le07} describing $T=1$ pairing, color pairing and $T=0,1$ pairing respectively.
For the case of $T=0,1$ we examine both exactly solvable sets, the degenerate orbital model and the RG model with non-degenerate single particle orbits, in turn.

\section{The multicomponent seniority Hamiltonian}
The pairing Hamiltonian with degenerate orbitals can be solved
analytically if the interaction has a separable structure with
pair operators.  The Hamiltonian is given by
\be
H = - G \sum^\Omega_{i,j} \sum_\alpha P^\dagger_{\alpha i}P_{\alpha j}.
\ee
Here $\Omega$ is the number of spatial orbitals and $P^\dagger$ is a pair
creation operator.  The label $\alpha$ indicates the
quantum numbers associated with the fermion internal degrees of freedom.
More specifically, for the nuclear physics application, we distinguish
isospin quantum numbers of the pairing operators and write
the Hamiltonian as

\begin{equation}
H= -G \sum_{ij}^\Omega \sum_\mu (P_{\mu
i}^{\dagger }P_{\mu j}+Q_{\mu i}^{\dagger }Q_{\mu j}),
\label{H0}
\end{equation}
with
\begin{equation}
P_{\mu i}^{\dagger }=\frac{1}{\sqrt{2}}\left\{ a_{i}^{\dagger
}a_{i}^{\dagger }\right\} _{0\mu }^{01};~Q_{\mu i}^{\dagger }=\frac{1}{\sqrt{%
2}}\left\{ a_{i}^{\dagger }a_{i}^{\dagger }\right\} _{\mu 0}^{10}.
\label{PQ}
\end{equation}%
In the expression for the pair creation operators $P^\dagger,Q^\dagger$,
the first column in the coupling refers to spin and the second to isospin.
The six pair creation and annihilation operators together with the spin operator, the isospin operator and the spin-isospin tensor close the $SO(8)$ algebra.
The Hamiltonian (\ref{H0}) with equal strength for the two pairings ($T=0$ and $T=1$) is invariant under $SU(4)$ transformations of the fermion
basis.

Moreover, the proton-neutron pairing Hamiltonian (\ref{H0}) is completely equivalent to a spin $S=3/2$ fermion pairing Hamiltonian appropriate to describe trapped cold atoms with four hyperfine components \cite{Cap07}. The pair operators that close the SO(8) algebra in this case are
\begin{equation*}
S_{i}^{\dagger }=\frac{1}{\sqrt{2}}\left\{ a_{i}^{\dagger
}a_{i}^{\dagger }\right\} _{0 }^{0};~D_{\mu i}^{\dagger }=\frac{1}{\sqrt{%
2}}\left\{ a_{i}^{\dagger }a_{i}^{\dagger }\right\} _{\mu}^{2},
\end{equation*}%
where the spin $3/2$ fermion pairs couple to total spin $S=0$ (monopole $S$ pair) and $S=2$ (quadrupole $D$ pair), and the $SU(4)$ symmetric Hamiltonian reads
\begin{equation}
H= -G \sum_{ij}^\Omega (S_{i}^{\dagger }S_{j}+\sum_{\mu=-2}^{2} D_{\mu i}^{\dagger }D_{\mu j}).
\label{H2}
\end{equation}

There is a one-to-one correspondence between Hamiltonians (\ref{H0}) and (\ref{H2}) if the pair operators are related as $Q_{\pm 1}^{\dagger }=D_{\pm 2}^{\dagger}$, $P_{\pm 1}^{\dagger }=D_{\pm 1}^{\dagger }$, $P_{0}^{\dagger }=D_{0}^{\dagger }$, and $Q_{0}^{\dagger }=S^{\dagger }$. Therefore, all conclusions we will extract from the proton-neutron $SU(4)$ symmetric pairing Hamiltonians are equally valid for the monopole and quadrupole pairing Hamiltonians of spin $3/2$ fermions.

Coming back to the Hamiltonian (\ref{H0}), we shall also consider the identical-particle pairing Hamiltonian
keeping only the two $P$-type operators $P^\dagger_{1i},P^\dagger_{-1i}$,
and the isovector pairing Hamiltonian keeping all three $P$-type operators.
To define a bulk limit
for Hamiltonian of the form Eq. (1), we let both the number of
spatial orbitals $\Omega$
and the number of particles $N$ go to infinity, but keeping the
fractional filling fixed at some value $f$.
The bulk limit also requires that the energy scale with the size
of the system.  This implies that
$G$ vary inversely to $\Omega$ for large values of $\Omega$,
i.e.
\be G={g\over \Omega},
\ee with $g$ a constant.
\subsection{Identical-particle pairing}
As a warm-up exercise, we begin with the well-known case of
the seniority Hamiltonian for the two-component fermion system,
eg. only neutrons or only protons.
The formula for the energy of the identical-particle seniority
Hamiltonian is \cite[Eq. 6.12]{RiSch}
\be
E_v(N_\tau) = -{1\over 4} G ( N_\tau - v)(2 \Omega -N_\tau -v +2 ),
\label{E0_ex}
\ee
where $v$ is the seniority and $N_\tau=N_p,N_n$ is the number of nucleons of
isospin $\tau$.  The ground state has seniority $v=0$.
In terms of $g$ and $f=N_\tau/2\Omega$, the energy per particle
can be expressed
\be
\label{E_0}
{E_0\over N_\tau} = -{1\over 2}g  \left(1 -f + {1\over \Omega}\right).
\ee
The corresponding expressions for the energy in the
BCS theory are\cite[Eq. 6.62]{RiSch}
\be
\label{E_BCS}
E_{BCS}=-{1\over 4} G N \left( 2 \Omega -N + {N\over \Omega}\right)
\ee
and
\be
{E_{BCS}\over N_\tau}=
-{1\over 2} g (1-f+f/\Omega ).
\ee
The energies in eq. (\ref{E_0}) and (\ref{E_BCS}) only differ by a
term of order  ${\cal O}(g/\Omega)$.  Thus, the
BCS theory for identical particles and the seniority Hamiltonian
is exact in the bulk limit.

We now ask about the accuracy of Eq. (\ref{E_BCS})  in the
more general context where both neutrons and protons are together.
For simplicity, we only consider the case of equal numbers of neutrons
and protons,
\be
N_p = N_n = {N\over 2}.
\ee
The naive extension of the BCS theory is to simply add the energies
of the neutron and proton condensates independently.  Then Eq. (\ref{E_BCS})
would still be valid.

\subsection{Isovector and $SU(4)$-symmetric Hamiltonians}
The corresponding formulas for isovector pairing
are given in ref. \cite[Eq. (9)]{ev81}.  The eigenvalues are
\be
E_{v,T}(N) = -{1\over 8}G(N-v) ( 4 \Omega -N -v + 6) + {G\over
2}T(T+1);\,\,\,\,\,
\ee
where $T$ is the isospin of
the eigenfunction.  Let us assume that $N_n$ and $N_p$ are even.
Then the ground state  has seniority $v=0$ and isospin $T=0$.  The energy per particle is
\be
{E_0\over N} = -{1\over 2}g\left(1-f + {3\over 2\Omega}\right).
\ee
One sees that the leading term is the same as in the identical-particle
Hamiltonian.  The finite-size correction is different, giving more binding
for isovector pairing.

The general formula for the energies of the $SU(4)$-symmetric Hamiltonian
is derived in ref. \cite{fl64}.  It is rather complicated for general
seniority, but for $v=0$ it reduces to a form very similar to Eq. (\ref{E_0})
\cite{ev81,en97},
\be
	E_{0,\lambda_2}(N) =
 -{1\over 8}GN ( 4 \Omega -N + 12) + {G\over2}\lambda_2(\lambda_2+4).
\ee
Here $\lambda_2$ is the spin-isospin label of of $SU(4)$.  In the ground state
$\lambda_2=0$ and  the energy per particle is
\be
{E_0\over N} = -{1\over 2}g\left(1-f + {3\over \Omega}\right).
\ee
Again the bulk limit is the same as in identical-particle pairing.  Note
also that the finite-size correction is larger than either of the two other
Hamiltonians.

\subsection{Quasiparticle and collective excitations}
  In the BCS theory of the identical-particle Hamiltonian, all excited
states are quasiparticle excitations, whose energies
can be calculated by blocking individual orbitals from participating in the
condensate.  The lowest excitation is a two-quasiparticle state;
its energy is obtained by substituting $\Omega \rightarrow \Omega -2 $
and $N \rightarrow N -2 $ in Eq.~(\ref{E_BCS}).
The resulting excitation energy
\be
\label{2q}
E_{2q} = g + {\cal O}(\Omega^{-1})
\ee
exhibits a finite gap in the excitation energy spectrum that persists to
the thermodynamic limit.  For the exact solutions, the two-quasiparticle
excitations are given by the states of seniority $v=2$. Formulas identical
to Eq.~(\ref{2q}) are obtained by taking differences of energies between
states of $v=2$ and $v=0$ and keeping the other quantum numbers fixed.

The other important manifestation of the BCS condensation is the
number-parity dependence of the
binding energies.  The systems with odd $N$ are described as
one-quasiparticle states in BCS and as states with seniority $v=1$ in the
exact solutions.  In all cases the change in energy adding one particle
is given by
\be
E_q  = E(N+1) - E(N)= {g\over 2} + {\cal O}(\Omega^{-1})\,\,\,(N {\rm~even})
\ee
$$
  = -{g\over 2} + {\cal O}(\Omega^{-1})\,\,\, (N {\rm~odd})
$$
The pairing gap $\Delta$ is conventionally defined as half the second-order
difference of binding energies, with a sign chosen to have $\Delta$ a
positive quantity.  The formula is
\be
\Delta_{o-e}(N) = {1\over 2} \left( 2 E(N) -E(N-1)-E(N+1)\right)
\ee
for $N$ odd.  Again, the BCS and the exact solutions all have the
same thermodynamic limit,
\be
\Delta_{o-e} = {g\over 2} + {\cal O}(\Omega^{-1}).
\ee

While the quasiparticle excitations are the same in BCS and the
exactly solvable Hamiltonians, the latter have
additional excitations corresponding to the other quantum numbers.
For nonzero values of $T$ in the isovector pairing Hamiltonian, the
lowest excitation has $v=0,T=2$ with an excitation energy of
\be
E_{0,2}-E_{0,0} = {3 g \over \Omega}.
\ee
This is inside the quasiparticle gap and goes to zero in the bulk limit.
The behavior of the $SU(4)$-invariant Hamiltonian is very similar;
the excitations with $v=0$ and finite
$\lambda_2$ are inside the quasiparticle gap and have vanishing
excitation energy in the bulk limit.

The collective excitations can be viewed as arising from the
degeneracy of the BCS solutions with respect to choice of the
paired orbitals.  For the isovector Hamiltonian, the pair
condensate wave function may be rotated in isospin space to give
a degenerate ground state; the wave function transforms as the
$[11]$ representation of the $SU(2)$ of isospin group.  Similarly,
for the $SU(4)$-symmetric Hamiltonian, the condensate transforms as
the $[11]$ representation of that group.   This is all contained
in the manifold of solutions of the Hartree-Fock-Bogoliubov equations.
These equations give the natural generalization of BCS to permit
arbitrary pair assignments in constructing the condensate.

\section{The $SO(8)$ Richardson-Gaudin  model}

To make a more realistic model for pairing in large systems, the Hamiltonian
needs to include a single-particle term as well as the interaction.
Richardson has given a construction of the exact many-body wave function
for two-component fermion systems with an arbitrary single-particle
Hamiltonian and the separable pairing form for the interaction \cite{Rich63}.
The combination of the Richardson solution with the integrable Gaudin magnet model \cite{Gau76} gave rise to a large family of fully integrable and exactly solvable models based on the $SU(2)$ algebra named Richardson-Gaudin models \cite{du04}.  More recently the RG models have been generalized to simple algebras of arbitrary rank. In particular, the $SO(8)$ RG models describe isoscalar-isovector pairing Hamiltonians with non-degenerate single particle orbits \cite{le07}. In the simplest version of the model the Hamiltonian is $SU(4)$-symmetric, however symmetry breaking terms can be added to the single particle Hamiltonian within the exact solution. The exact solution of the $SO(8)$ model is obtained by solving four sets of algebraic nonlinear coupled equations in terms of four sets of unknown spectral parameters. Each independent solution of the coupled set of equations completely determines an eigenstate and the corresponding eigenvalue. One of the sets of spectral parameters is composed of the pair energies that fix the structure of the correlated pairs in the single particle basis as in the $SU(2)$ case. The other three sets of spectral parameters are responsible for the structure of the eigenstates in the space of the internal degrees of freedom (spin-isospin).
The thermodynamic limit of the exact Richardson $SU(2)$ solution has been obtained making use of
an exact mapping between the quantum many-body problem and a classical electrostatic problem in two dimensions \cite{Rom02,Ort05}. This mapping is difficult to pursue for higher rank RG models due to the proliferation of different species of charges (spectral parameters). Instead, we will extrapolate to the bulk limit the numerical exact results that can be obtained for quite large systems.

The $SU(4)$ symmetric RG pairing Hamiltonian that we consider is
\be
\label{SU4-sp}
H=\sum_{i}^{\Omega} {\varepsilon}_i \hat N_{i}-
\frac{g}{\Omega}\sum_{ij\mu }^{\Omega}(P_{\mu
i}^{\dagger }P_{\mu j}+D_{\mu i}^{\dagger }D_{\mu j}),
\ee
where $\hat N_i$ is the number operator counting all particles in
level $i$ and the isovector $P_{\mu i}^{\dagger }$ and isoscalar  $Q_{\mu i}^{\dagger }$ are defined in (\ref{PQ}).
The single particle energies are taken equally spaced as $\varepsilon_i={1\over 2 \Omega }(i-1)$ with $i$ an integer in the interval $[1,\Omega]$.
 We carried out the calculation for quarter filling ($N=\Omega$)
and interaction strength $g=0.15$, considering systems of different sizes
\cite{g}. For these system parameters the fraction of the condensate is $\approx 0.54$. This value is estimated by counting the ratio of complex pair energies from the exact solution to the total number of pairs. As discussed in  \cite{Ort05} a complex pair energy implies the formation of a correlated Cooper pair, while the rest of pair energies, being real, describe free fermions.

\subsection{The BCS limit }
In the thermodynamic limit the  BCS equations for the Hamiltonian
Eq. (\ref{SU4-sp}) reduce to%
\begin{equation*}
4\int_{0}^{1/2}{\left( {1-\frac{{\varepsilon -\lambda }}{\sqrt{\Delta
^{2}+(\varepsilon-\lambda )^{2}}}}\right) }d\varepsilon =1
\end{equation*}%
\begin{equation*}
\int_{0}^{1/2}{\frac{1}{\sqrt{\Delta ^{2}+(\varepsilon -\lambda )^{2}}}}%
d\varepsilon =\frac{1}{g},
\end{equation*}

and for the BCS energy we have :

\begin{equation*}
\frac{E_{BCS}}{N}=4\int_{0}^{1/2}{\left( {1-\frac{{\varepsilon -\lambda }}{%
\sqrt{\Delta ^{2}+(\varepsilon -\lambda )^{2}}}}\right) }\varepsilon
d\varepsilon -\frac{2}{g}\Delta ^{2}.
\end{equation*}

The solution for the given parameter values has chemical potential
$\lambda =  0.12468144   $ and gap parameter $\Delta =  0.015466976   $.    The quasiparticle energy of level $q$ in BCS is given by
\be
E_q = \sqrt{(\varepsilon_q - \lambda)^2+\Delta^2}   + \lambda.
\ee

\subsection{Thermodynamic limit of the exact solution}
As before, we want to compare the exact solution to the BCS approximation in the
thermodynamic limit, $N,\Omega\rightarrow \infty $ with $N/\Omega$ constant.
We first focus on the ground-state energy per-particle.
We assume it is linear in
$N$ in the thermodynamic limit, and we expand $E(N)/N$ as
\be
\label{expansion}
{E\over N}=a+b/N+c/N^2+d/N^3+{\cal O}\left( 1/N^{4}\right).
\ee
For our analysis we find the exact numerical solutions for $N$ in
the range $160 \le N \le 1000$ and fit the parameters of the expansion Eq.
(\ref{expansion}).  For details of how we solve the equations for
the $SO(8)$ RG model, see ref. \cite{le07}.
The results are shown in Table I and Fig. 1. We see that the fit with a cubic polynomial
reproduces the BCS energy with seven significant figures.
\begin{table}
\caption{Ground-state energy, quasi-particle energy, and different gaps as defined in the text for the $SU(4)$-symmetric pairing Hamiltonian in the
${1/N}$ cubic  expansion, with a comparison to bulk BCS limit.}
\begin{tabular}{|c|c|c|c|c|c|}
\hline
             &Method  & $a$        & $b$         & $c$ & $d$\\ \hline\hline
$\frac{E}{N}$ & Exact  & $0.062022149$ & $-0.597581$ & $1.278831$ & $-11.1571$\\
             & BCS    & $0.062022154$ &    &  &\\ \hline
$E_q$ & Exact & $0.140148$ & $-0.479740 $ & $-10.0327$  &$-1107.25$ \\
     & BCS    & $0.140151$ &            &  &     \\ \hline
$\Delta_{o-e}$ & Exact    & $0.0154637$    & $-0.699890 $   & $-2.24642$  & $-1066.63 $ \\
             &  BCS      &  $0.0154669$    &     &    &  \\ \hline
$\Delta_c$ & Exact & $0.0154672$ & $0.0961964$ & $2.59458$  & $ -257.910$\\
           &  BCS  & $0.0154669$ &            &   & \\ \hline
\end{tabular}
\end{table}
\begin{figure}
\includegraphics [width = 11cm]{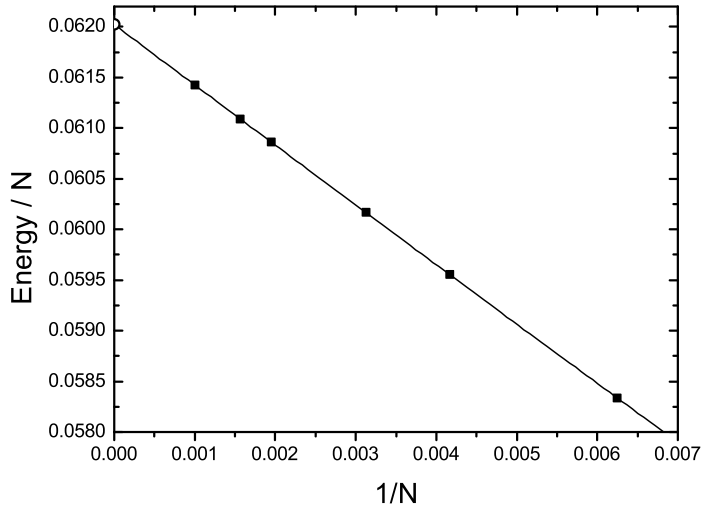}
\caption{\label{energy} Ground-sate energy per-particle of the $SU(4)$-symmetric fermion systems
as a function of particle number $N$.  Black squares show the calculated
results and the solid line is the cubic fit of Eq. (\ref{expansion}). The open circle is the bulk BCS limit.
}
\end{figure}
We next examine the
quasiparticle energy $E_q$, calculated as the binding difference between
the systems with $4n$ and $4n+1$ particles,
\be
E_q(4n) =E(4n+1) -E(4n)\approx a + b/N + c/N^2+d/N^3,
\ee
We show in the
Table the $1/N$ expansion for this quantity.
In our case we choose the blocked orbital as the lowest unoccupied level corresponding to $\varepsilon_q=1/8$.
We expect that the constant term will be the same for addition and removal
energies, but the higher order terms in the expansion will differ.

We also report in the Table the pairing gap computed in two different ways.
The first is the binding energy difference $\Delta_{o-e}$
center at particle number $N=4n+1$,
\be
\Delta_{o-e}(4n+1) = {1 \over 2} \left[ 2 E(4n+1) - E(4n) - E(4n+2) \right].
\label{del0}
\ee
Another, more theoretical definition is the canonical gap,
\begin{equation}
\Delta_c =\frac{1}{8}\frac{g}{\Omega}\sum\limits_{i=1,\sigma \tau}^{\Omega}\sqrt{n_{i\sigma \tau}(1-n_{i\sigma \tau})},
\label{delc}
\end{equation}
where $n_{i\sigma \tau}$ is the occupation probability of a fermion with spin $\sigma$ and isospin $\tau$ in level $i$. Note that this definition coincides with the BCS gap if the occupation probabilities are calculated in the BCS approximation.

The cubic fits to the quasiparticle energy and the odd-even and canonical gaps are shown in Table 1. The precision with which the thermodynamic limit of the exact solution approaches the BCS results for these quantities is slightly lower than that of the ground state energy. The reason is that the quasiparticle energy and the odd-even gap result from the difference of big numbers (ground state energies of finite systems) while the canonical gap involves a further calculation of the occupation numbers once the solution of the spectral parameters is obtained. In any case the precision obtained, five or  six significant figures, is enough to guarantee that the thermodynamic limit of the exact RG model coincides with the bulk BCS values for the four magnitudes studied.
The convergence of the two gaps with the size of the system is shown in Fig. 2. One can see that for these
quantities the expansion is rapidly convergent as for the ground state energy, and that it can be reliably
extrapolated. A noteworthy fact is that both definitions of the gap, the odd-even energy difference (\ref{del0}) and the canonical gap (\ref{delc}), converge to the bulk BCS gap, in spite of the fact that the approaching direction of convergence is opposite.

\begin{figure}
\includegraphics [width = 11cm]{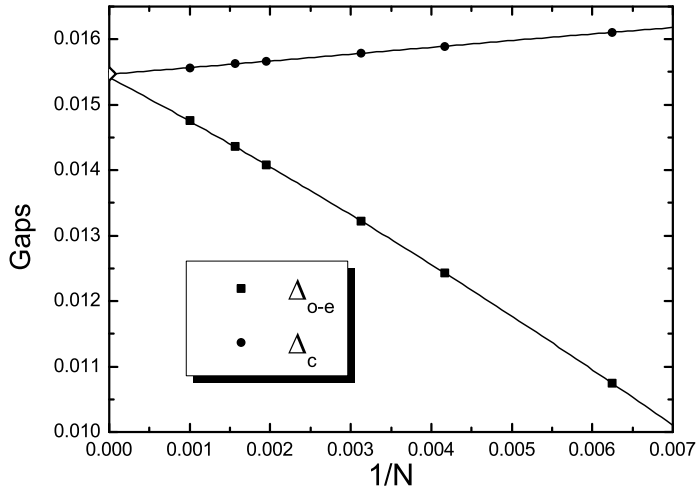}
\caption{\label{gaps} Gaps of the $SU(4)$-symmetric fermion systems
as a function of particle number $N$. The continuous lines are the cubic fits. The empty diamonds is the bulk BCS gap parameter.
}
\end{figure}

\section{Perspective}

   The excellent agreement between the BCS and the extrapolated exact solutions
of our numerical example shows that higher order correlations
on top of a single pair condensate are not important for the bulk ground state characteristics of
many-component fermion systems, when the Hamiltonian is of the pairing
type given in Eq. (\ref{SU4-sp}).  However, the interaction in that
Hamiltonian is still rather restrictive.
For example, one could imagine a two-particle interaction of the 4-component
system that supported a 4-particle bound state (the alpha particle in
nuclear physics) but no heavier ones.  Then
the many-body system in the Bose-Einstein condensate limit would be a gas of the 4-particle clusters.
Analogously, in the weak coupling regime there might be stronger quartet correlations than in the current RG models.  It
might be interesting to ask what conditions would be needed to give the
two-particle interaction a capability to produce extensive higher-order
correlation effects.

  Although for pairing Hamiltonians like (\ref{SU4-sp}) the higher correlations do not affect the bulk gaps and
per-particle condensation energy, there are still collective excitations
present in the system than are beyond the reach of the BCS approximation.
This is seen most clearly
in the analytic formulas of the pure seniority Hamiltonians discussed
in Sect. II.  It might be interesting to see how well the standard
methods of many-body theory perform for calculating the effects of
the collective degrees of freedom starting from the HFB solutions.
For example, collective effects are often treated in nuclear theory
by the Quasi-Particle Random Phase Approximation or by the
Generator Coordinate Method.  It fact, it has been shown already
that for the 2-component fermion system the particle-particle RPA and QRPA give very accurate
corrections to BCS for the total ground-state energy \cite{ba74,du03} of finite systems in the fluctuation dominated regime and in the superconducting regime respectively. However, close to the phase transition region both approaches cannot reproduce the large quantum fluctuations. It would be interesting to carry out these analysis in the thermodynamic limit, as well as to consider quantum corrections beyond BCS to multicomponent fermion systems having the exact solution of the higher rank RG models as a benchmark reference.

GFB and JD acknowledges discussions with N. Sandulescu.
This study arose out of discussions at a workshop at the Institute
for Nuclear Theory.  We thank the INT for its support under DOE
grant DE-FG02-00ER41132.   
This work was supported in part by
 grant FIS2009-07277 of the Spanish Ministry of Science and Innovation.


\begin{thebibliography}{100}
\bibitem{RiSch} ``The Nuclear Many-Body Problem,'' P. Ring, and P. Schuck,
(Springer,  2004).
\bibitem{Petry85} H. R. Petry, et al., Phys. Lett. B{\bf159}, 363 (1985).
\bibitem{Bohr09} H. Bohr, and J. da Providencia, J. Phys. A: Math. Theor. {\bf41}, 405202 (2008).
\bibitem{fl64} B. H. Flowers and S. Szikowski, Proc. Phys. Soc. {\bf84}, 673 (1964).
\bibitem{ev81} J. A. Evans, et al. Nucl. Phys. A {\bf367}, 77 (1981).
\bibitem{en97} J. Engel, et al., Phys. Rev. C {\bf55}, 1781 (1997). D. R. Bes, et al. Phys. Rev. C {\bf61}, 024315 (2000).
\bibitem{du04} J. Dukelsky, S. Pittel, and G. Sierra, Rev. Mod. Phys. {\bf76}, 643 (2004).
\bibitem{duk06} J. Dukelsky, et al. Phys. Rev. Lett. {\bf96}, 072503 (2006).
\bibitem{duk09} B. Errea, J. Dukelsky, and G. Ortiz, Phys. Rev. A {\bf79}, 051603 (2009).
\bibitem{le07} S. Lerma H., B. Errea, J. Dukelsky, and W. Satula,
Phys. Rev. Lett {\bf 99}, 032501 (2007).
\bibitem{Cap07} S. Capponi, et al. Phys. Rev. B {\bf75}, 100503(R) (2007).
\bibitem{Rich63} R. W. Richardson, Phys. Lett. {\bf3}, 277 (1963).
\bibitem{Gau76} M. Gaudin, J. Phys. (Paris) {\bf37}, 1087 (1976).
\bibitem{Rom02} J. M. Roman, G. Sierra, and J. Dukelsky, Nucl. Phys. B {\bf634}, 483 (2002).
\bibitem{Ort05} G. Ortiz, and J. Dukelsky, Phys. Rev. A {\bf72}, 043611 (2005).
\bibitem{g} Phenomenological values for $g$ in nuclear physics are in the
range $g=0.2-0.3$, depending the energy cutoff in the orbital space.

\bibitem{ba74} J. Bang and J. Krumlinde, Nucl. Phys. {\bf A141}, 18 (1970).
\bibitem{du03} J. Dukelsky, G. Dussel, J. Hirsch and P. Schuck, Nucl.
Phys. A {\bf 714}, 63 (2003).


\end{thebibliography}
\end{document}